\pdfoutput=1

\documentclass{article}
\usepackage[utf8]{inputenc}
\usepackage{graphicx}
\usepackage[colorlinks,allcolors=cyan!70!black]{hyperref}

\usepackage{amsmath}
\usepackage{authblk}
\usepackage{bm}
\usepackage{acronym}
\usepackage{cleveref}
\usepackage{mathtools}
\usepackage{siunitx}
\usepackage{xspace}

\usepackage{xcolor}

\usepackage{natbib}
\setcitestyle{numbers,square,comma}

\newcommand{\PARAM}{\bm{\vartheta}}

\newcommand{\HPARAM}{\bm{\psi}}

\newcommand{\OBSS}{{\bm{y}}}

\newcommand{\DATUM}{{d}}
\newcommand{\DATA}{{\bm{\DATUM}}}

\DeclareDocumentCommand \pr { m o } { \IfNoValueTF{#2} {p(#1)}{p(#1 \,|\, #2)} }

\newcommand{\FVK}{\mathrm{FvK}}

\newcommand{\meanv}{0.94}

\newacro{BASIS}{Bayesian annealed sequential important sampling}
\newacro{DAG}{directed acyclic graph}
\newacro{DPD}{dissipative particle dynamics}
\newacro{MAP}{maximum a posteriori}
\newacro{ML}{maximum likelihood}
\newacro{RBC}{red blood cell}
\newacro{SFS}{stress free state}
\newacro{TMCMC}{transitional Markov chain Monte-Carlo}
\newacro{TTF}{tank treading frequency}
\newacro{Ca}{Capillary number}
\newacro{NN}{neural network}
\newacro{SDE}{stomatocyte-discocyte-echinocyte}

\newcommand{\mir}{Mirheo\xspace}
\newcommand{\SM}{supplementary material\xspace}
\newcommand{\tRBC}{{t-RBC}\xspace}

\title{The stress-free state of human erythrocytes: data driven inference of a transferable RBC model}

\author[1,2]{Lucas Amoudruz}
\author[1,2]{Athena Economides}
\author[1,2]{Georgios Arampatzis}
\author[1,2]{Petros Koumoutsakos\thanks{petros@seas.harvard.edu}}

\affil[1]{Computational Science and Engineering Laboratory, ETH Z\"{u}rich, CH-8092, Switzerland.}
\affil[2]{School of Engineering and Applied Sciences, Harvard University, Cambridge, MA 02138, United States.}

\begin{document}

\maketitle

\begin{abstract}
 The stress-free state (SFS) of red blood cells (RBCs) is a fundamental reference configuration for the calibration of computational models, yet it  remains unknown.
 Current experimental methods cannot measure the SFS of cells without affecting their mechanical properties while computational postulates are the subject of controversial discussions.
 Here, we introduce data driven estimates of the SFS shape and the visco-elastic properties of RBCs.
 We employ data from single-cell experiments that include measurements of the equilibrium shape, of stretched cells, and relaxation times of initially stretched RBCs. A hierarchical Bayesian model accounts for these experimental and data heterogeneities.
 We quantify, for the first time, the SFS of RBCs and use it to introduce a transferable RBC (\tRBC) model.
 The effectiveness of the proposed model is shown on predictions of unseen experimental conditions during the inference, including the critical stress of transitions between tumbling and tank-treading cells in shear flow.
 Our findings demonstrate that the proposed \tRBC model provides predictions of blood flows with unprecedented accuracy and quantified uncertainties.
\end{abstract}


\section*{Introduction}

\Acp{RBC} are vital elements of blood as they are responsible for the delivery of oxygen to the entire human body. As they traverse the microcirculature \acp{RBC} undergo highly non-linear deformations, that are accommodated by their visco-elastic properties~\citep{caro2011}.
These properties are mainly controlled by the structure of  their membrane, composed of a lipid bilayer anchored on a network of proteins (cytoskeleton), and enclosing a viscous solvent (hemoglobin).
The \ac{RBC} membrane and the hemoglobin are both considered incompressible.
The cytoskeleton and the lipid bilayer of the membrane provide elastic resistance against local shearing, stretching, and bending.
Over the last two decades numerous mathematical models for the \ac{RBC} membrane have been proposed, aiming to explain complex phenomena, and complement experimental studies through parametric exploration and system optimization~\citep{freund2014}. 

State-of-the-art models of \acp{RBC} account for shear deformation of the membrane with respect to a state at which the membrane has zero in-plane elastic energy, namely the \ac{SFS}~\citep{lim2008,khairy2008,cordasco2014,peng2014,peng2015,mauer2018}.
The existence of a non-spherical \ac{SFS} was demonstrated by the experimental results of \citet{fischer2004}, who showed that the \ac{RBC} membrane exhibits shape memory, and of \citet{dupire2012}, who suggested that shape memory can explain certain dynamical transitions of cells in shear flow.
\Citet{vsvelc2012} suggested an analysis to compare the deformation of the cytoskeleton in a micropipette for a given \ac{SFS} shape to that measured by experiments~\citep{lee1999}, but did not infer the \ac{SFS} from the experimental data.
Furthermore, current experimental methods do not allow to directly measure the \ac{SFS} of cells without affecting their mechanical properties (see section 2.3.3 of \citet{lim2008} and references therein).
For these reasons,  previous works have performed parametric studies using a predefined \ac{SFS}. Such calibrations affect the dynamics of \acp{RBC}, and in turn are key factors when comparing  computational and experimental data~\citep{levant2016a, cordasco2014, peng2014, tsubota2014a, lim2008}.
An ever increasing amount of evidence from both experiments and simulations have shown that the \ac{SFS} of the membrane skeleton is neither the biconcave resting shape, nor a spherical shell~\citep{cordasco2014, peng2014, tsubota2014a}.
The consensus on the \ac{SFS} is an oblate-like shell, with the same surface area and a larger volume than the RBC, though the exact \ac{SFS} remains elusive~\citep{levant2016a,cordasco2014,peng2014}.

Several computational studies have performed parametric investigations to quantify the effect of the \ac{SFS} on the response of \acp{RBC}, under static and dynamic conditions~\citep{li2005, lim2008, tsubota2014a, cordasco2014, peng2014, cordasco2017}.
The \ac{SFS} calibration shape was shown to significantly affect predictions of the \ac{RBC} dynamics in simple shear flow.
In particular, computational findings~\citep{cordasco2014, peng2014} demonstrated that the \ac{SFS} alters not only the critical shear rate separating tumbling and tank-treading \ac{RBC} dynamics, but also the motion of the \ac{RBC} membrane at the critical shear rate.
\citet{peng2014} searched for a family of \acp{SFS}, which could reproduce at the same time the biconcave resting shape and the dynamics of single erythrocytes in simple shear flow.
Their findings show that a \ac{SFS} closer to a sphere, rather than to a biconcave disk, approaches not only the experimental critical shear stress, but also preserves the experimentally observed biconcave shape during tank-treading~\citep{dupire2012}. This finding was in contrast to previous computational studies.
As a result, \citet{peng2014} envisioned that \ac{RBC} dynamics at low shear rates might enable the quantification of the \ac{SFS}.

We complement the aforementioned studies, by performing a data-driven inference of the  \ac{SFS} and its potential variability in the population of healthy \acp{RBC}.
We use hierarchical Bayesian inference to integrate data from multiple experimental sources and conditions, and generate a data-informed probabilistic \ac{RBC} model that incorporates modeling and experimental uncertainties in its predictions~\citep{economides2021}.
The structure of the model incorporates the variability~\citep{reichel2019} of \acp{RBC} elastic properties.
In contrast to the recent study by \citet{economides2021}, where the \ac{RBC} model was considered a ``black-box'' with an arbitrary, predefined \ac{SFS}, here we perform a global sensitivity analysis for each quantity of interest.
Inert factors are excluded during the inference process, to reduce the computational cost and avoid numerical artifacts while sampling the posterior distribution~\citep{arampatzis2018langevin}.
The high computational cost associated with the Bayesian inference is mitigated by the use of \acp{NN} as emulators of the \ac{RBC} model output.
This approach enables the simultaneous inference of all material properties in the employed \ac{RBC} model.
In particular, the \ac{SFS} (parameterized by its reduced volume), shear (both linear and non-linear components), and bending moduli are inferred from experimental data of \acp{RBC} in equilibrium~\citep{evans1972a}, and under stretching~\citep{mills2004, suresh2005}.
In turn, the membrane viscosity is inferred from experiments of \ac{RBC} relaxation after elongation~\citep{Hochmuth1979}.

Predictions of the fully-calibrated model are validated against experimental data coming from complex flow conditions that were not part of the inference.
Specifically, the calibrated model captures the velocity and elongation of \acp{RBC} flowing in a microtube~\citep{tomaiuolo2009}, the \ac{TTF} and inclination angle of \acp{RBC} in simple shear flows~\citep{fischer2015}, and, most importantly, the critical shear stress between the tumbling and tank-treading motion of \acp{RBC} in shear flow~\citep{fischer2013}.
Our findings demonstrate, for the first time, the transferability of the inferred model, without problem specific tuning, and its capability to predict complex flow configurations that were not part of the inference.

\section*{Methods}

\subsection*{Red Blood Cell model}

We model the \ac{RBC} membrane as a surface whose dynamics evolve according to bending resistance of the lipid-bilayer, shear and dilation elasticity of the cytoskeleton  and membrane viscosity. The shear and dilation elasticity are minimal at the \ac{SFS} of the \ac{RBC}, a state that is not known.
The resistance to bending is described by the energy
\begin{equation} \label{eq:energy:bending}
  U_{bending} = 2 \kappa_b \oint{H^2dA},
\end{equation}
where the integral is taken over the membrane, $\kappa_b$ is the bending modulus and $H$ is the mean curvature of the membrane.
The in-plane elastic energy accounts for the shear and dilation elasticity of the cytoskeleton,
\begin{equation} \label{eq:energy:inplane}
  U_{in-plane} = \frac {K_\alpha}{2} \oint {\left( \alpha^2 + a_3 \alpha^3 + a_4 \alpha^4 \right) dA_0} + \mu \oint{ \left( \beta + b_1 \alpha \beta + b_2 \beta^2 \right) dA_0},
\end{equation}
where the integral is taken over the \ac{SFS} surface, $\alpha$ and $\beta$ are the local dilation and shear strain invariants of the membrane, respectively, $K_\alpha$ is the dilation elastic modulus, $\mu$ is the shear elastic modulus and the coefficients $a_3$, $a_4$, $b_1$ and $b_2$ are parameters that control the non-linearity of the membrane elasticity for large deformations~\citep{lim2008}.

The membrane is discretized into a triangle mesh composed of $N_v$ vertices with positions $\mathbf{r}_i$, velocities $\mathbf{v}_i$ and mass $m$, $i=1,2,\dots,N_v$, evolving according to Newton's law of motion.
The bending energy described by \cref{eq:energy:bending} is discretized following~\citet{julicher1996,bian2020a} and the in-plane energy is computed as described in \citet{lim2008}.
The forces arising from these energy terms are formed by the negative gradient of the energy with respect to the particle positions.
The membrane viscosity is modeled through pairwise forces between particles sharing an edge in the triangle mesh.
The viscous force exerted by particle $j$ to particle $i$ is given by~\citep{fedosov2010pHD}
\begin{equation}
  \mathbf{f}_{ij}^{visc} = - \gamma \left(\mathbf{v}_{ij} \cdot \mathbf{e}_{ij} \right) \mathbf{e}_{ij},
\end{equation}
where $\gamma$ is the friction coefficient, $\mathbf{v}_{ij} = \mathbf{v}_i - \mathbf{v}_j$ and $\mathbf{e}_{ij}$ is the unit vector between $\mathbf{r}_i$ to $\mathbf{r}_j$.
The membrane viscosity depends linearly on the friction coefficient, $\eta_m = \gamma \sqrt{3} / 4$.
Finally, the constraints of preserving the area of the membrane and the volume of the cytosol are enforced through energy penalization terms,
\begin{equation*}
  U_{area} = k_A \frac{\left(A - A_0\right)^2}{A_0}, \quad
  U_{volume} = k_V \frac{\left(V - V_0\right)^2}{V_0},
\end{equation*}
where $A_0$ and $V_0$ are the area and volume of the cell at rest and $A$ and $V$ are the area and volume of the cell, respectively.
The coefficients $k_A$ and $k_V$ are chosen empirically with values that are large enough to enforce the conservation of the membrane area and volume of the \ac{RBC}.
More details on the discretization of the energies are presented in the \SM.

\begin{figure}
  \centering
  \includegraphics[width=0.8\textwidth]{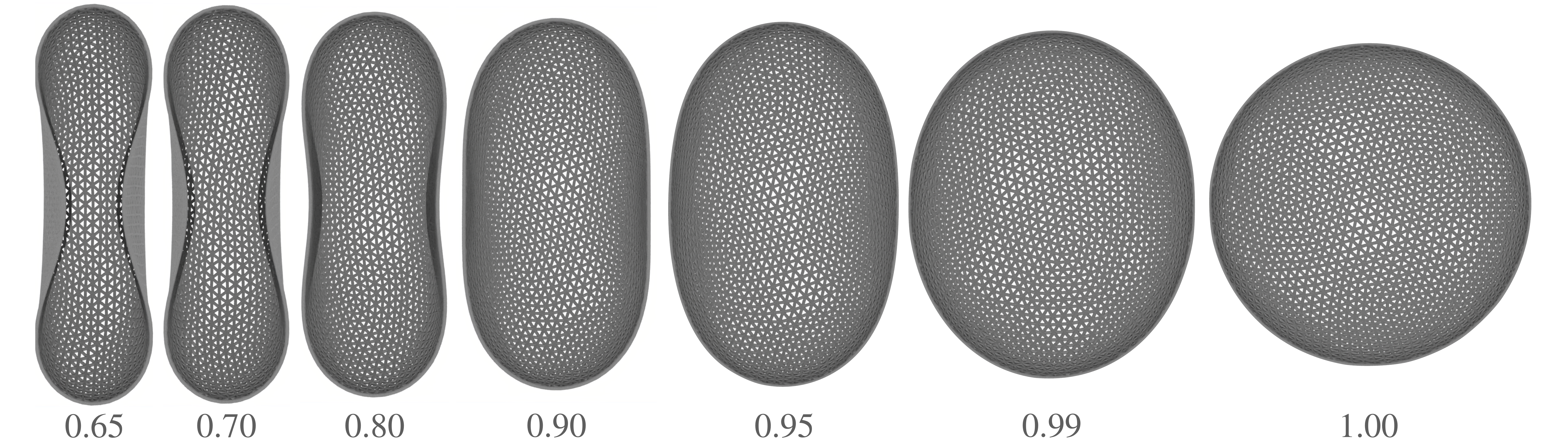}
  \caption{\Ac{SFS} shapes of different reduced volumes $v$ (indicated below each shape).
    All shapes are axi-symmetric around the horizontal axis.
    \label{fig:SFS:sequence}}
\end{figure}

The \ac{SFS} of the \ac{RBC} is parameterized by its reduced volume $v$, i.e., the volume of the \ac{SFS} relative to that of a sphere with same area as the \ac{SFS}.
Following \citet{lim2008}, the \ac{SFS} is obtained by minimizing the energy of a membrane with bending resistance, shear and dilation elastic energy with a sphere as reference state.
The area of the \ac{SFS} is constrained to that of a healthy \ac{RBC} and the volume is parameterized by the reduced volume $v$ (ratio of the volume with respect to that of a sphere with the same area).
With $v$ ranging from 0.65 to unity, this procedure results in biconcave shapes, oblates and spheroids at low, intermediate and high reduced volume, respectively (\cref{fig:SFS:sequence}). We note that the value of this reduced volume is chosen arbitrarily in \ac{RBC} models and accordingly affects their dynamics.

The parameters governing the \ac{RBC} mechanics comprise the reduced volume of the \ac{SFS} $v$, the shear modulus $\mu$, the shear-hardening coefficient $b_2$, the bending modulus $\kappa_b$ and the membrane viscosity $\eta_m$.
These parameters are calibrated from experimental data sets that we introduce in the next sections.
The remaining parameters of the model are chosen as follows:
the dilation elastic modulus is set to $K_\alpha = \mu$;
the non-linear coefficients in the shear energy formulation are set to $a_3 = -1$, $a_4 = 8$ and $b_1 = 0.7$~\citep{lim2008};
the area and volume of the cells are fixed to $A_0=\SI{135}{\micro\meter^2}$ and $V_0=\SI{94}{\micro\meter^3}$, respectively~\citep{evans1972a}.

\subsection*{Heterogeneous data and a probabilistic model for the RBC: \tRBC}

We link seven experimental data sets measured from three experimental conditions with the computational model using a hierarchical statistical framework.
The first data set corresponds to the measurements of the diameter $D$, maximal thickness $h_{max}$ and minimal thickness $h_{min}$ of single cells at equilibrium, as reported by \citet{evans1972a}.
The second and third data sets are measurements of the two principal diameters of \acp{RBC} stretched by two micro-beads.
The micro-beads are attached to the membrane at two opposite sides of the cell's rim and are pulled by forces of magnitude $F_{ext}$ in opposite directions.
The two largest principal diameters of the cells, $D_{ax}$ and $D_{tr}$, are reported by \citet{mills2004} and \citet{suresh2005} against the stretching force magnitude $F_{ext}$.
The remaining data sets were collected by \citet{Hochmuth1979} from initially stretched \acp{RBC} relaxing to their equilibrium shape.
The data sets consist in the ratio of the two principal diameters of the cells, $D_{ax} / D_{tr}$ measured at constant time intervals.

\begin{figure}
  \centering
  \includegraphics[width=0.7\textwidth]{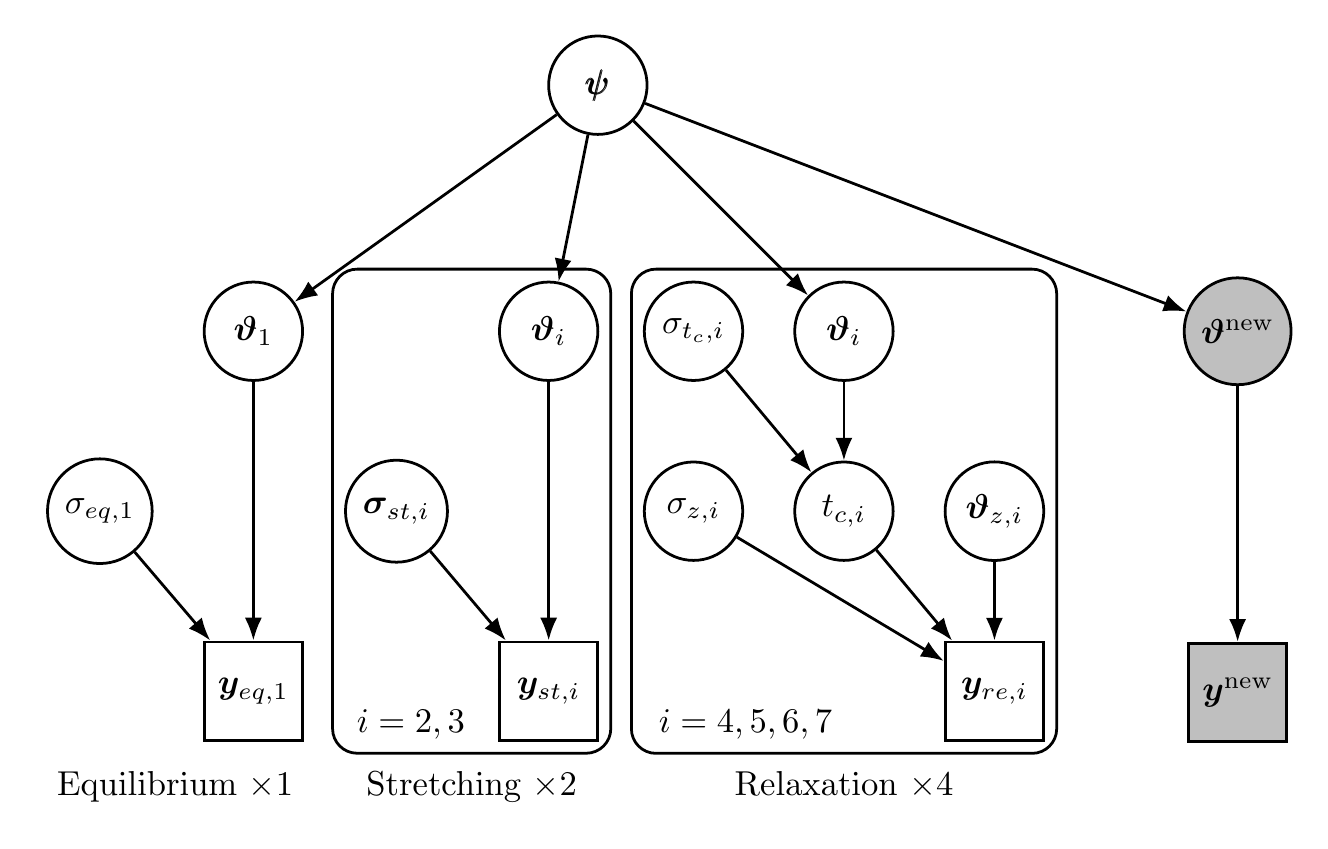}
  \caption{Structure of the \tRBC model, presented as a DAG.
    Rectangular and circular nodes are observed and unobserved quantities, respectively.
    The arrows represent the causal links between variables.
    Rounded rectangles are repeated depending on the number of data sets for each case.
    Shaded nodes are not part of the inference and are used to predict configurations that were not used during the inference phase.
    \label{fig:DAG:hierarchical}}
\end{figure}

We assume that each data set is one realization of the random variable $\OBSS_{\alpha,i}$ (called observable), where $\alpha$ denotes the experimental conditions (equilibrium, stretching or relaxation) and $i$ is the index of the data set (we drop the indices in the remaining of this section to lighten the notations).
The \tRBC model relates the computational model and its parameters to the probability distribution of the observable.
The structure of the \tRBC model, represented as a \ac{DAG}, is shown on \cref{fig:DAG:hierarchical}.
We distinguish the parameters of the computational model, $\PARAM = \left( v, \mu, \kappa_b, b_2, \eta_m \right)$, from those of the error model (explained below) such as the standard deviation $\sigma$.
In addition, we introduce the hyper-parameter $\HPARAM$ that is further discussed below.

The hierarchical structure of the \tRBC model represents two levels of uncertainty.
First, the computational parameters $\PARAM_i$ for each data set $i$ is drawn from a distribution parameterized by the hyper parameter $\HPARAM$, $\pr{\PARAM_i}[\HPARAM]$, representing the variability of the cells properties.
This variability is due to the origin of the cells (from different donors), the age of the cells and the different experimental conditions.
Second, for each data set, the observable is assumed to be normally distributed around the output of the computational model.
This second level of uncertainty reflects the measurement errors and the inaccuracy of the computational model.
The measurements errors are modeled separately for each experimental conditions, with parameters $\sigma$ as shown in \cref{fig:DAG:hierarchical}.
We note that in the case of the relaxation experiment, we introduced an intermediate variable, $t_c$, which is the relaxation time of the cell.
This addition simplifies the inference procedure as the initial shape of the \ac{RBC} in experiments is unknown.
Instead, we assume that $t_c$ depends on the \ac{RBC} parameters only and is independent on the initial shape of the cell.
This assumption allows to estimate $t_c$ from the computational model with an arbitrary initial stretched shape.
The data is then modeled as an exponential decay with rate $t_c^{-1}$ and additional parameters contained in $\PARAM_z$.
The exact dependencies between the random variables are detailed in the \SM.

\subsection*{Offline surrogate of the computational model}

The evaluation of the computational model for each experimental condition (cell equilibration, stretching and relaxation), while relatively fast thanks to the high-performance implementation in \mir~\citep{alexeev2020a}, remains computationally costly  for performing Bayesian inference of the \tRBC model presented above.
Instead, we replace the computational model during the Bayesian inference with an offline surrogate.

\begin{figure}
  \centering
  \includegraphics[width=\textwidth]{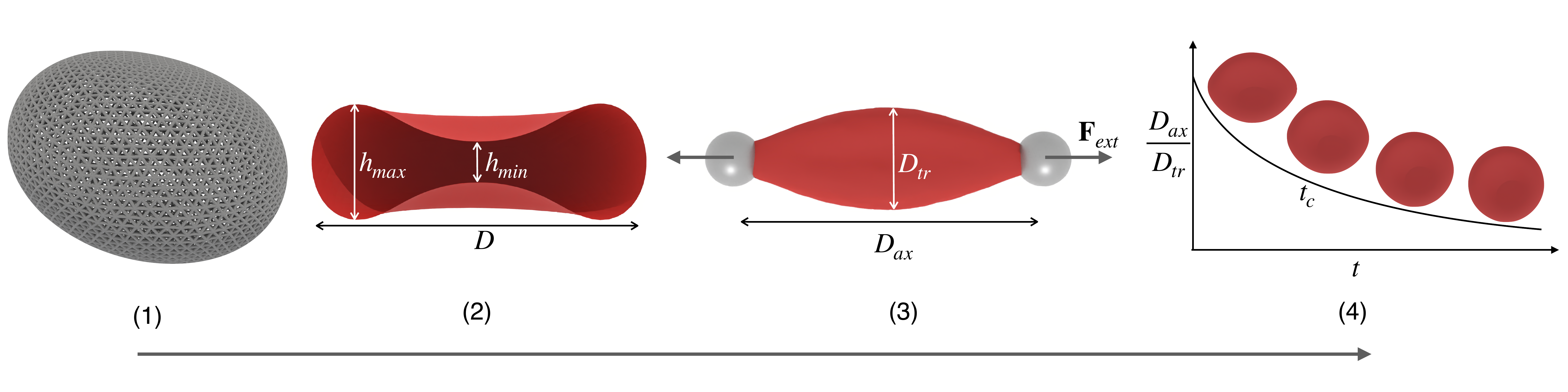}
  \caption{Sequence of simulations used to compute the output of all experiments for a set of parameters $\PARAM$.
    (1) Generation of the \ac{SFS} mesh, needed by all subsequent simulations.
    (2) Generation of the equilibrium shape.
    (3) Stretching of the equilibrated cell.
    (4) Relaxation of the stretched cell.
    \label{fig:experiments}}
\end{figure}

The surrogate is formed by three \acp{NN}, one for each experimental condition, that takes as input the computational parameters $\PARAM$ (and the stretching force magnitude $F_{ext}$ for the stretching case) and outputs the observable of the computational model ($(D, h_{min}, h_{max})$ for the equilibration case, $(D_{ax}, D_{tr})$ for the stretching case and $t_c$ for the relaxation case).
Each \ac{NN} is composed of three hidden layers of 32 neurons and hyperbolic tangent activation gates.
The training data was generated for 50'000 samples uniformly distributed in the input space of the surrogate.
The corresponding observable values were then computed with the procedure described in \cref{fig:experiments} using the computational model.
The \ac{NN} parameters were then trained on these samples (split into 80\% and 20\% training and validation sets, respectively) to minimize the mean squared error between the \acp{NN} and the computational model.
The training was performed with the Adam optimizer and we used early stopping to avoid over-fitting.
The prediction accuracy of the surrogate is shown in the \SM.

\section*{Results and Discussion}

\subsection*{Bayesian Inference}

We infer the parameters of the \ac{RBC} model from the combined experimental data sets using hierarchical Bayesian inference.
The posterior distribution of the parameters is sampled using \ac{BASIS}~\citep{wu2017a}, an unbiased version of \ac{TMCMC}~\citep{ching2007}.
This sampling method does not rely on the gradient of the model with respect to the parameters.
In this situation, regions of constant likelihood lead to poor sampling~\citep{raue2013}. In turn, we eliminate  the parameters that are  inert for the respective experimental condition.
We performed a sensitivity analysis of the model output with respect to the parameters in the \SM.
The results indicate that the combination of the three experimental cases chosen in this study are complementary for the inference of the cell parameters: the equilibrium shape is sensitive to $v$ and $\FVK$ (where $\FVK = \mu A_0 / 4\pi \kappa_b$ is the F\"oppl-von K\'arm\'ann number), the stretched cell diameters vary mainly with $\mu$ and $b_2$ (and $v$ and $\FVK$ at low stretching forces) and the relaxation characteristic time is sensitive to $\mu$ and $\eta_m$.

\begin{figure}
  \centering
  \includegraphics[width=\textwidth]{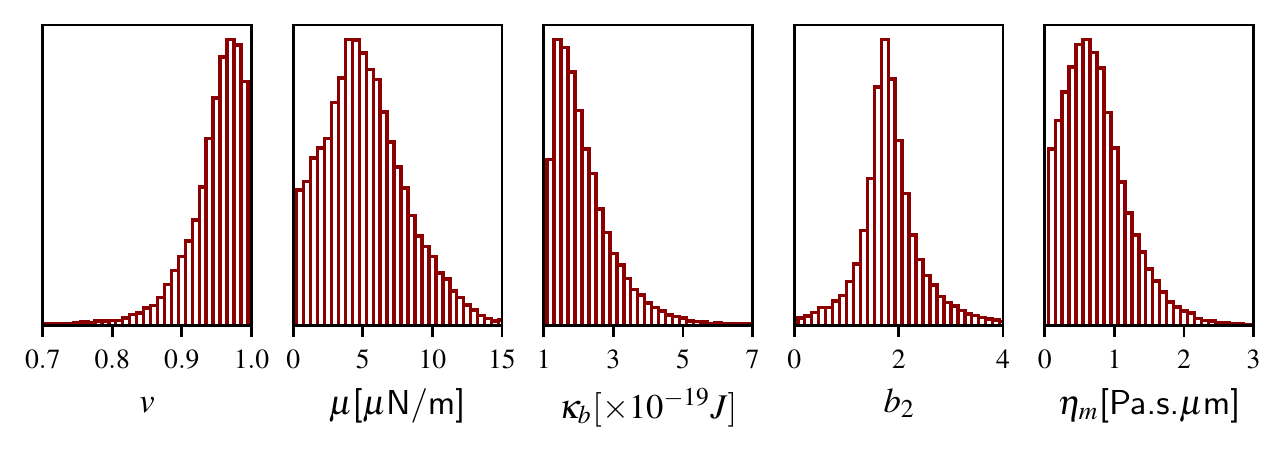}
  \caption{Posterior distribution of the \ac{RBC} parameters $\pr{\PARAM^\text{new}}[\DATA]$.
    Only the marginal distributions are shown since the variables are independent.
    \label{fig:thetanew}
  }
\end{figure}

The parameters of the \tRBC model are sampled as described in the \SM using the Korali framework~\citep{martin2021a}.
The resulting posterior distribution of the \ac{RBC} parameters, $\pr{\PARAM^\text{new}}[\DATA]$, is shown on \cref{fig:thetanew}, with corresponding mean, median, \ac{ML} and \ac{MAP} values reported in \cref{tab:stats:thetanew}.
All distributions have a clear peak with relatively high uncertainties around the \ac{MAP}, due to the heterogeneity of the data sets.

The inferred shear modulus has a mean at $\mu=\SI{4.99}{\micro\newton\per\meter}$, which is within the range of values used in previous studies: $\SI{6.3}{\micro\newton\per\meter}$~\citep{fedosov2010a,fedosov2010b}, $\SI{2.42}{\micro\newton\per\meter}$~\citep{turlier2016}, $\SI{4.5}{\micro\newton\per\meter}$~\citep{yazdani2016}.
Similarly, the inferred bending modulus is consistent with the values used in previous works ($\kappa_b = \SI{2.4e-19}{\joule}$~\citep{fedosov2010a}, $\SI{4.8e-19}{\joule}$~\citep{fedosov2010b}, $\SI{1.43e-19}{\joule}$~\citep{turlier2016}, $\SI{3.0e-19}{\joule}$~\citep{yazdani2016}).
In addition, the inferred membrane viscosity is close to that found in \citet{walchli2020a} ($\SI{0.63}{\pascal\second\micro\meter}$) and in \citet{Hochmuth1979} ($0.6-\SI{0.8}{\pascal\second\micro\meter}$).
The parameter $b_2$ is found higher than in \citet{lim2008} ($b_2 = 0.75$).
However, the cell deformations were most likely smaller in the latter study than in the cell-stretching experiments that were used for the inference.

The inferred reduced volume of the \ac{SFS} has a mean around $v=\meanv$, which suggests that the \ac{SFS} is more likely an oblate than the biconcave shape, based on these experimental data sets.
This value is close to those used in previous studies ($v=0.95$~\citep{khairy2008}, $v=0.96$~\citep{mauer2018}, $0.90 \leq v \leq 0.998$~\citep{peng2014,peng2015}, $v=0.997$~\citep{cordasco2014}).
Furthermore, the range of values obtained from the Bayesian inference agrees with the conclusions of \citet{lim2008}, who showed that $0.925 \leq v \leq 0.976$ to reproduce the \ac{SDE} sequence observed when changing the bending properties of the lipid-bilayer of the membrane.
Similarly, \citet{geekiyanage2019} concluded that the reduced volume of the \ac{SFS} is around $v=0.94$ to obtain the \ac{SDE} sequence.
We note that the result of \citet{lim2008} was obtained with predefined values of the mechanical properties of the membranes, while in the current work the mechanical properties of the membranes are inferred together with the \ac{SFS} reduced volume.
Furthermore, the studies that inferred the \ac{SFS} reduced volume based on the \ac{SDE} sequence did not consider the dynamics of the \acp{RBC} in dynamic flow conditions~\citep{lim2008, geekiyanage2019},

\begin{table}
  \centering
  \begin{tabular}{lccccc}
    & mean & median & ML & MAP & standard deviation \\
    \hline
    $v$        & \meanv & 0.95 & 0.96 & 0.96 & 0.04 \\
    $\mu$      & 4.99 & 4.68 & 4.60 & 4.60 & 2.24 \\
    $\kappa_b$ & 2.10 & 1.85 & 1.46 & 1.46 & 0.93 \\
    $b_2$      & 1.84 & 1.73 & 1.69 & 1.69 & 0.82 \\
    $\eta_m$   & 0.69 & 0.62 & 0.66 & 0.66 & 0.46
  \end{tabular}
  \caption{Statistics on the posterior distribution of the parameters based on all the experimental data sets.
    The parameters $\mu$, $\kappa_b$ and $\eta_m$ are expressed in $\SI{}{\micro\newton\per\meter}$, $\SI{1e-19}{\joule}$ and $\SI{}{\pascal\second\micro\meter}$, respectively.
    \label{tab:stats:thetanew}
  }
\end{table}

\begin{figure}
  \centering
  \includegraphics[width=0.8\textwidth]{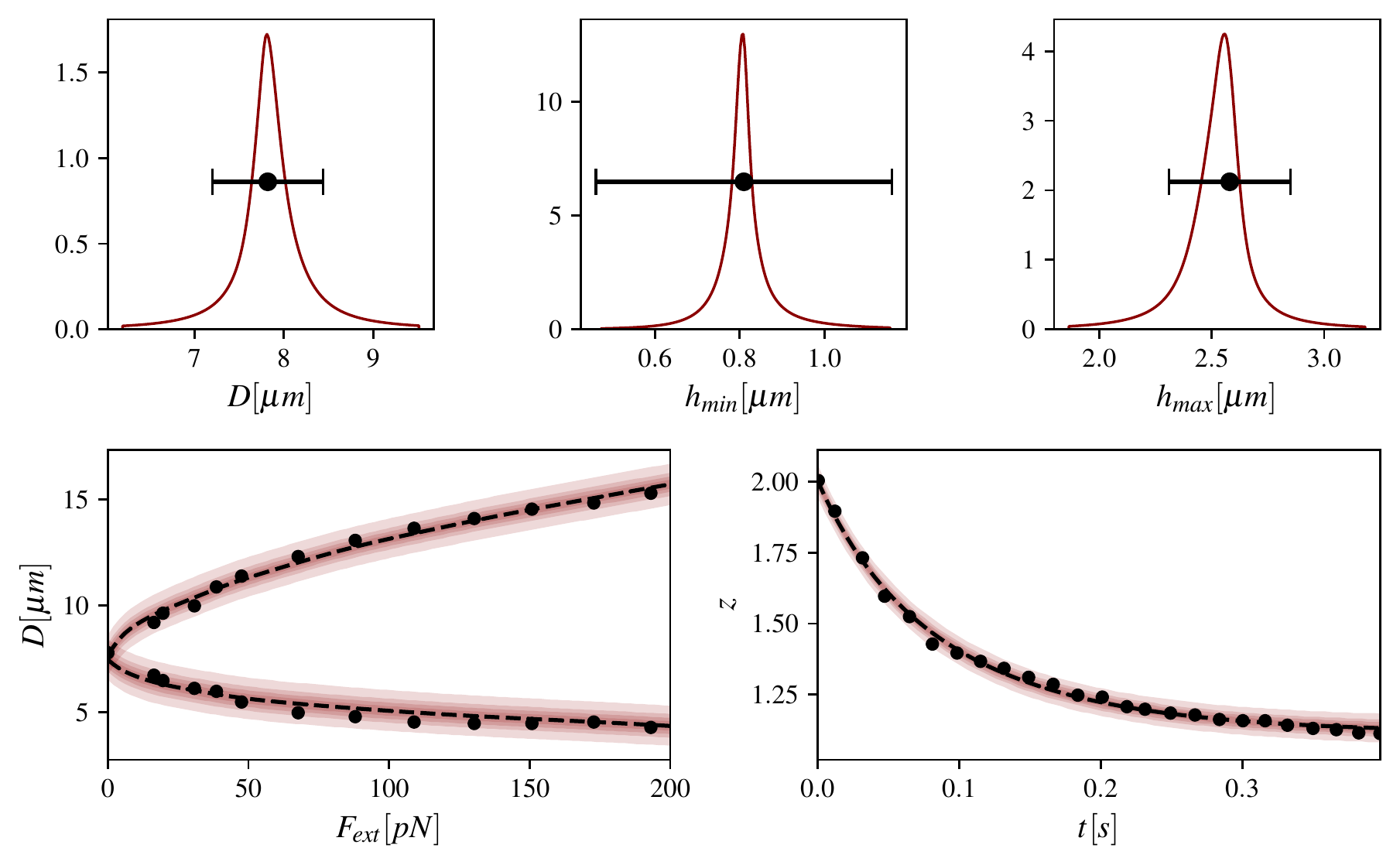}
  \caption{Forward predictions of the \tRBC model on the single-cell experiments.
    Top: Probability distribution of the diameter (left), minimal thickness (middle) and maximal thickness (right) of an equilibrated cell.
    The symbols and error bars denote the measurements and corresponding standard deviations reported by \citet{evans1972a}, respectively.
    Bottom left: Cell diameters against the stretching force magnitude. Mean prediction (dashed line) and experimental data from \citet{mills2004} (symbols). The shaded regions denote the 50\%, 75\%, 90\% and 99\% credible intervals of the predictions.
    Bottom right: Ratio of the cell diameters $z=D_{ax}/D_{tr}$ against time of an initially stretched RBC.
    The symbols are the experimental data from \citet{Hochmuth1979}.
    \label{fig:forward}}
\end{figure}

The inferred parameters are then tested against the experimental data sets used for the inference.
\Cref{fig:forward} shows the predictions of the \tRBC model for one data set for each experimental condition.
The parameters used for the predictions are sampled from the probability distribution $\pr{\PARAM_i}[\DATA]$, where $\DATA$ contains the seven data sets used for the inference.
In all cases, the experimental data lie inside the credible intervals given by the \tRBC model.

\subsection*{Model generalization}
\label{se:predictions}

Contrary to the one-at-a-time approach, a commonly used practice for validating \ac{RBC} models~\citep{Kotsalos2019, Dupin2007, fedosov2010a}, we test the predictive accuracy of the calibrated \ac{RBC} model in configurations that were not seen during the inference.
The posterior distribution of the parameters was inferred using simple experimental conditions where only one or two parameters had a significant effect on the output in each case.
Here the calibrated model is validated in complex dynamic situations, where multiple parameters affect the output, as shown in parametric studies found in the literature (details below).
In particular, we test the model prediction on five quantities: the \ac{TTF}, inclination angle, and threshold shear stress for tumbling-to-tank-treading transition of \acp{RBC} in simple shear flow, the elongation of \acp{RBC} flowing through a microtube and their respective velocity against the applied pressure gradient.
The \ac{TTF} and inclination angle are known to be significantly affected by the membrane viscosity~\citep{Yazdani2013}.
The threshold shear stress for tumbling-to-tank-treading transition depends on the \ac{SFS}~\citep{cordasco2014, peng2014}, and the length of flowing \acp{RBC} in microtubes depends on the bending stiffness of the membrane~\citep{noguchi2005}.
We emphasize that it is crucial to estimate the prediction capabilities of the model on data coming from conditions not seen during the inference phase to test the transferability of the model.

The following cases are substantially more expensive in terms of computations than those used to calibrate the model (each evaluation takes at least 24 hours on a single P100 graphics processing unit).
Therefore, instead of propagating the posterior distribution of the parameters through the computational model, we evaluate each quantity of interest with the mean estimates of the posterior distribution.

\subsubsection*{RBC in a circular microtube}

\begin{figure}
  \centering
  \includegraphics[width=0.49\textwidth]{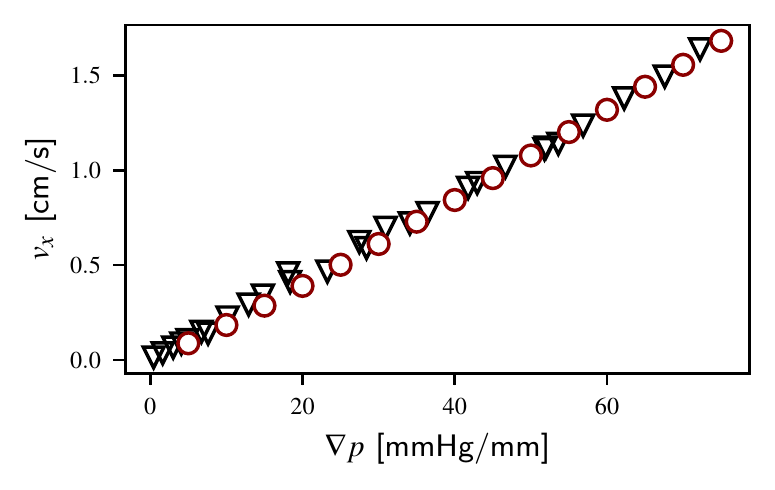}
  \includegraphics[width=0.49\textwidth]{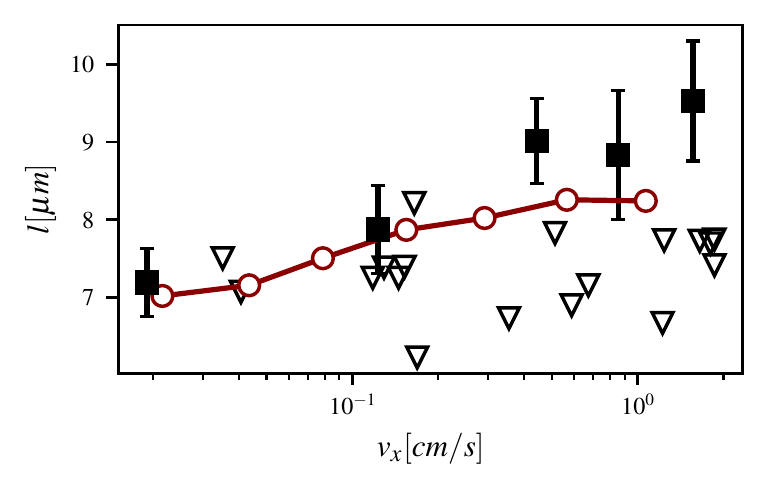}
  \caption{\Ac{RBC} flowing in a straight circular microtube of radius $R=\SI{3.3}{\micro\meter}$.
    Left: Velocity of the cell $v_x$ against the applied pressure gradient $\nabla p$.
    Experimental data from \citet{tomaiuolo2009} (triangles) and simulation results (circles).
    Right: Length of the \ac{RBC} $l$ against the velocity $v_x$.
    Experimental data from \citet{tomaiuolo2009} and \citet{hochmuth1970} (squares and triangles, respectively) and simulation results (open circles).
    \label{fig:microtube}
  }
\end{figure}

Single \acp{RBC} flowing in straight microtubes adopt a steady parachute-like shape.
The cell length $l$ and velocity $v_x$ depend on the flow rate and the radius $R$ of the tube.
A pressure difference $\Delta p$ between the ends of the tube causes the solvent and the cell to flow.
The tube has a length $L \gg R$ large enough so that the cells reach an equilibrium shape before the measurements.
The length and velocity of the cells, $l$ and $v_x$, were recorded for different pressure gradients $\nabla p = \Delta p / L$ experimentally for $R=\SI{3.30}{\micro\meter}$~\citep{tomaiuolo2009} and $R=\SI{3.35}{\micro\meter}$~\citep{hochmuth1970}.
Simulations of this system are performed with the current calibrated model (using the mean of the posterior distributions) for $R=\SI{3.30}{\micro\meter}$ (see \SM for details).
The simulations show a good agreement with the experimental data (\cref{fig:microtube}).
The variability of the cell lengths in the experiments could be attributed to the variability in the mechanical properties of the cells, but also to that of the cell sizes.

\subsubsection*{RBC in a linear shear flow}

Single \acp{RBC} suspended in a linear shear flow exhibit rich dynamics.
At low shear rates, the cells tumble (rotate in a rigid-like motion).
Increasing the shear rate above a threshold value causes the cell membrane to tank tread: the cell adopts an elongated shape forming an angle $\theta$ with the flow direction, while the membrane rotates around the cell with a frequency $f$ (the \ac{TTF}).
Below we present predictions of the \tRBC model for the inclination angle, the \ac{TTF} and the critical shear stress for tumbling to tank-treading transition.
These predictions were obtained for a fixed cytosol viscosity (see \SM).
However, we remark that this quantity is known to depend on the hemoglobin concentration, which varies notably with the age of the cells, and it may be of importance to model this variation in further research~\citep{chien1970}.

\paragraph{Inclination angle.}

\begin{figure}
  \centering
  \includegraphics[width=0.49\textwidth]{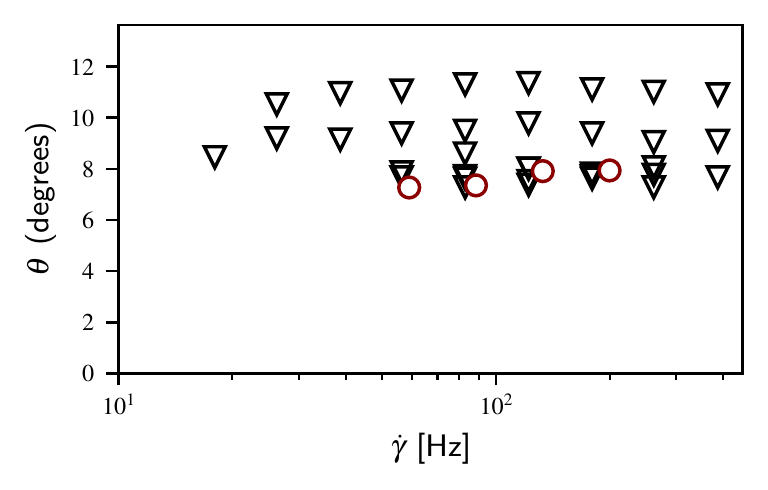}
  \includegraphics[width=0.49\textwidth]{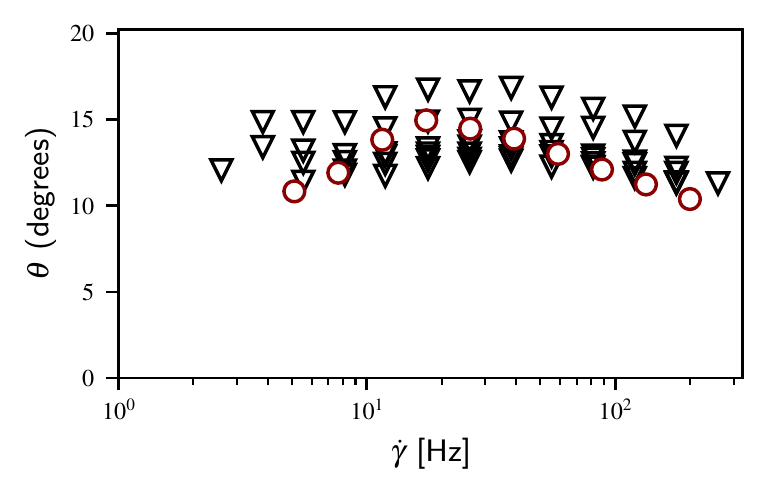}
  \caption{Mean inclination angle $\theta$ of tank-treading \acp{RBC} in a linear shear flow against the shear rate $\dot{\gamma}$.
    The triangles are data from \citet{fischer2015} and the empty circles are the simulation predictions obtained with the mean parameters of the posterior distributions.
    The left and right figures correspond to solvent viscosities $\eta = \SI{10.7}{\milli\pascal\second}$ and $\eta = \SI{23.9}{\milli\pascal\second}$, respectively.
  }
  \label{fig:TT:angle}
\end{figure}

Measurements of inclination angles of tank-treading \acp{RBC} in simple shear flows have been reported by \citet{fischer2015}.
The inclination angle $\theta$ obtained with the calibrated \ac{RBC} model (with the mean estimate of the parameters) is shown against the shear rate $\dot{\gamma}$ on \cref{fig:TT:angle}.
The model predictions are within the values observed experimentally.
In particular, for a solvent viscosity $\eta = \SI{23.9}{\milli\pascal\second}$, the model captures the trend of the experimental data, i.e. an increase of $\theta$ with $\dot{\gamma}$ followed by a decrease of $\theta$ above a critical shear rate.
This trend is less pronounced at the lower solvent viscosity $\eta = \SI{10.7}{\milli\pascal\second}$, both experimentally and in the simulations.

\paragraph{Tank treading frequency.}

\begin{figure}
  \centering
  \includegraphics[width=0.49\textwidth]{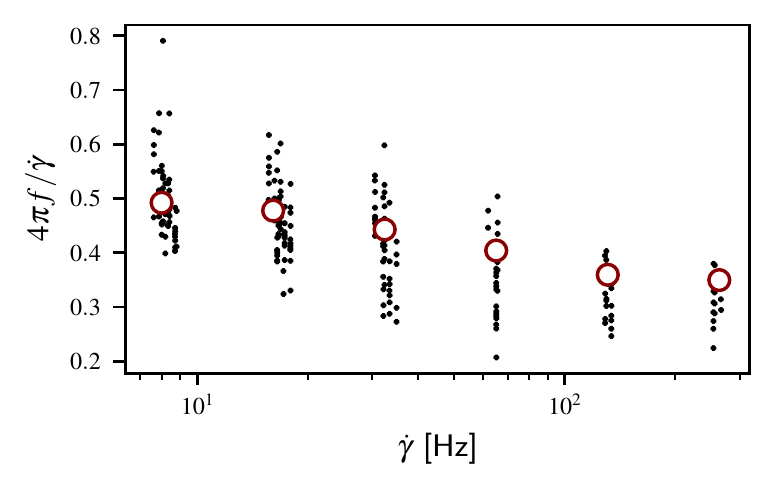}
  \includegraphics[width=0.49\textwidth]{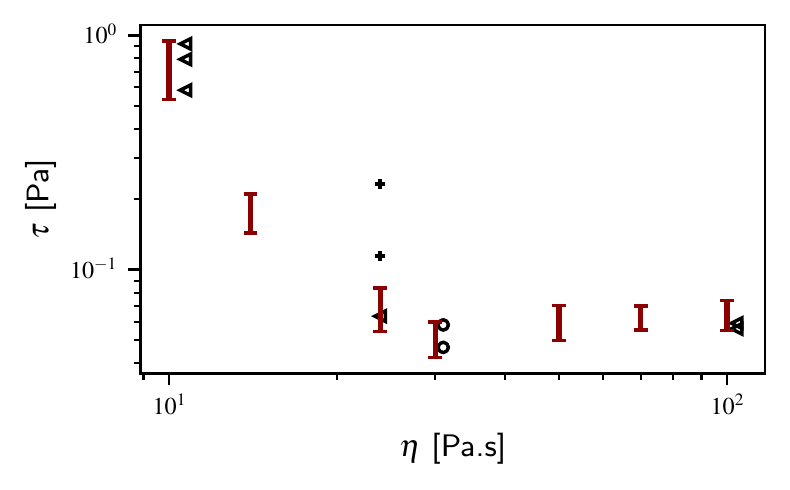}
  \caption{Left: \Ac{TTF} (normalized by the angular frequency of a sphere in a shear flow) of the \ac{RBC} in a linear shear flow against the shear rate $\dot{\gamma}$, with a solvent viscosity $\eta = \SI{28.9}{\milli\pascal\second}$.
    Experimental data from \citet{fischer2007} (dots) and simulation results with the mean parameters (open circles).
    Right: Critical shear stress $\tau = \dot{\gamma}\eta$ of the tumbling to tank-treading transition against the solvent viscosity $\eta$.
    Experimental data from \citet{fischer2013} (triangles and crosses) and \citet{abkarian2007} (circles), simulation results (bars).
    The crosses correspond to cells that underwent shape transitions due to chemicals.
    \label{fig:TTF}
  }
\end{figure}

\Cref{fig:TTF} shows the dimensionless \ac{TTF}, $4\pi f/\dot{\gamma}$, of a tank-treading \ac{RBC} suspended in a linear shear flow (with solvent viscosity $\eta = \SI{28.9}{\milli\pascal\second}$) for various shear rates $\dot{\gamma}$.
The \tRBC model predictions are performed with the mean estimate of the parameters.
Despite the complex dependency of the \ac{TTF} on the computational parameters, the \tRBC model shows a good agreement with the \ac{TTF} experimental data reported by \citet{fischer2007}.
Furthermore, the biconcave shape of the cell is preserved in the tank-treading simulations, as observed experimentally \citep{dupire2012} (see fig. S2).
However, we note that the biconcavity of the cell during tank treading is only reported qualitatively in experiments, and quantitative experimental data could help improving further the calibration of the model.

\paragraph{Tumbling to tank-treading transitions in linear shear flow.}

\Acp{RBC} in a linear shear flow undergo different regimes depending on the shear rate $\dot{\gamma}$ and solvent viscosity~\citep{yazdani2011}.
At large shear rates, the cell orientation oscillates around a steady angle while the membrane rolls, or ``tank-treads'' around the cell.
In contrast, \acp{RBC} rotate as a rigid object, or ``tumble'', when the shear rate is below a critical value.
The critical shear stress $\tau = \dot{\gamma} \eta$ has been measured experimentally by \citet{abkarian2007} and \citet{fischer2013} for different solvent viscosity $\eta$.
We performed numerical simulations with the calibrated model at different shear rates for several viscosity values $\eta$.
For a given solvent viscosity, the flow regime of the cell (tumbling or tank-treading) was reported for several shear rates.
The highest and lowest shear rates at which the \ac{RBC} tumbles and tank-treads, respectively, are reported on \cref{fig:TTF}.
The model predictions are in good agreement with experimental data.
As in the experimental data, the critical shear stress decreases sharply for solvent viscosity below $\SI{30}{\milli\pascal\second}$ and reaches a plateau above that viscosity.
Note that we show only the transitions for the discocytes in the data from \citet{fischer2013}.
The data points marked with crosses at $\eta = \SI{23.9}{\milli\pascal\second}$ are from cells that previously underwent shape transitions, possibly modifying their mechanical properties.
This observation probably explains the deviations between the shear stress obtained from experiments and that obtained from the simulations at this particular viscosity.

\subsection*{Limitations}

To minimize the computational cost of inference, we have fixed several parameters to specific values.
Specifically, the membrane model assumes zero spontaneous curvature (\cref{eq:energy:bending}), which typically results from differences in monolayer compositions found in-vivo.
However, discocytes, as shown in Figure 2.45 of \citet{lim2008}, have been found to exhibit a spontaneous curvature close to zero and we therefore ignored the spontaneous curvature in the current work.
Similarly, the area and volume of the \acp{RBC} exhibit a distribution that reflects the variability and aging of the cells.
While it is possible to include these variations in the statistical model, doing so would increase the complexity of the model and the computational cost for inference.
For similar reasons, we fixed the ratio $K_\alpha / \mu$ as in \citet{economides2021}.
While this choice is arbitrary, the current model predicts accurately various flow conditions.
The value of these parameters should ideally be inferred from additional experimental data and can be the subject of future research.

\section*{Conclusion}

We introduce a transferable \ac{RBC} model (\tRBC) that quantifies the visco-elastic membrane parameters and the \ac{SFS} of healthy \acp{RBC} through Bayesian inference.
The \tRBC model takes into account the cell heterogeneity, the measurement errors and the computational model inaccuracies.
The model parameters were calibrated on 7 data sets comprising measurements of \ac{RBC} dimensions at equilibrium, \ac{RBC} elongation under stretching forces and \ac{RBC} relaxation time.
The posterior distribution of the parameters have a relatively large standard deviation that possibly reflects the variability of mechanical properties among \acp{RBC}.
The reduced volume of the \ac{SFS} takes values that suggest that the cytoskeleton of \acp{RBC}, in its unstressed state, has an oblate shape.
The calibrated shear modulus, bending modulus and viscosity of the membrane were found to be in good agreement with previous studies, and we provide uncertainty on these parameters.

The calibrated model predicts accurately complex, single-cell dynamics, and agrees well with experimental data that were not used during the inference phase.
In particular, the calibrated model predicts accurately the velocity and length of cells flowing in narrow tubes, the inclination angle and \ac{TTF} of tank-treading cells in linear shear flows, and the critical shear stress of the tumbling to tank-treading motion of \acp{RBC} in linear shear flows.
We emphasize that the aforementioned quantities highly depend on the visco-elastic properties of the \ac{RBC} model, as demonstrated in numerous parametric studies in the literature.
The transferability of the proposed \tRBC model makes it a candidate of choice for predicting the dynamics of \acp{RBC} in previously unseen flow configurations that involve large deformations and/or complex dynamics.
In addition, the variability of the inferred parameters can be used to provide a more realistic description of blood flows with many cells, each cell having parameters drawn from the posterior density.
This approach would model the heterogeneity of the cells in blood.
The samples from the posterior density of the parameters are available online, together with the code used to produce the results of this study~\citep{tRBC_2022}.

\section*{Author Contributions}

AE, LA, PK designed the research.
LA, AE ran the \ac{RBC} simulations.
LA, AE, GA performed the Bayesian inference.
LA built the surrogate model.
LA, AE, PK interpreted the results.
LA, AE, GA, PK wrote the article.

\section*{Acknowledgments}

We would like to thank Xin Bian and Sergey Litvinov for their invaluable insights on \ac{RBC} modeling and simulations.
We acknowledge support by the The European High Performance Computing Joint Undertaking (EuroHPC) Grant DComEX (956201-H2020-JTI-EuroHPC-2019-1), and computational resources granted by the Swiss National Supercomputing Center (CSCS) under the project ID “s929”.

\section*{Declaration of Interest}

The authors declare no competing interests.

\bibliography{bibliography}



\end{document}